# Deep Learning (DL)-based Automatic Segmentation of the Internal Pudendal Artery (IPA) for Reduction of Erectile Dysfunction in Definitive Radiotherapy of Localized Prostate Cancer

Anjali Balagopal, Michael Dohopolski, Young Suk Kwon, Steven Montalvo, Howard Morgan, Ti Bai, Dan Nguyen, Xiao Liang, Xinran Zhong, Mu-Han Lin*, Neil Desai*, Steve Jiang*

Medical Artificial Intelligence and Automation (MAIA) Laboratory and Department of Radiation Oncology, University of Texas Southwestern Medical Center, Dallas, Texas, USA

*Co-correspondence authors. Emails: Mu-Han.Lin@UTSouthwestern.edu, Neil.Desai@UTSouthwestern.edu, Steve.Jiang@utsouthwestern.edu

## ABSTRACT

**Background and purpose:** Radiation-induced erectile dysfunction (RiED) is commonly seen in prostate cancer patients. Clinical trials have been developed in multiple institutions to investigate whether dose-sparing to the internal-pudendal-arteries (IPA) will improve retention of sexual potency. The IPA is usually not considered a conventional organ-at-risk (OAR) due to segmentation difficulty. In this work, we propose a deep learning (DL)-based auto-segmentation model for the IPA that utilizes CT and MRI or CT alone as the input image modality to accommodate variation in clinical practice.

**Materials and methods:** 86 patients with CT and MRI images and noisy IPA labels were recruited in this study. We split the data into 42/14/30 for model training, testing, and a clinical observer study, respectively. There were three major innovations in this model: 1) we designed an architecture with squeeze-and-excite blocks and modality attention for effective feature extraction and production of accurate segmentation, 2) a novel loss function was used for training the model effectively with noisy labels, and 3) modality dropout strategy was used for making the model capable of segmentation in the absence of MRI.

**Results:** The DSC, ASD, and HD95 values for the test dataset were 62.2±7.7%, 2.54±.87 mm, and 7±2.3mm, respectively. AI segmented contours were dosimetrically equivalent to the expert physician's contours. The observer study showed that expert physicians scored AI contours (mean=3.7) higher than inexperienced physicians' contours (mean=3.1). When inexperienced physicians started with AI contours, the score improved to 3.7.

**Conclusion:** The proposed model achieved good quality IPA contours to improve uniformity of segmentation and to facilitate introduction of standardized IPA segmentation into clinical trials and practice.

# I. INTRODUCTION

Treatment of localized prostate cancer (PCa) with radiotherapy (RT) is highly effective but incurs significant quality of life impact, of which erectile dysfunction (ED) is the most common long-term toxicity for initially potent men. Prior strategies to reduce ED after RT by dose-reduction to erectile tissue below the prostate [1] or with use of oral phosphodiesterase 5 inhibiors (i.e. RTOG 08-15)[2] have been unsuccessful.

Highly conformal RT techniques, in particular stereotactic ablative radiotherapy (SAbR), and magnetic resonance (MR) aided target and organ delineation in PCa have spurred renewed effort to spare critical neurovascular structures immediately adjacent to the prostate itself . In particular, a recent trial [3] suggested benefit to reducing dose to the internal pudendal arteries (IPA), the primary penile blood supply. On this basis, a multi-institutional randomized clinical trial led by our institution is investigating the benefit of individualized sparing of neurovascular structures, including the IPAs.  Because the IPA is not a commonly segmented structure, its standardized delineation was anticipated to be difficult and heterogeneously interpreted, challenging study quality. While we thus utilized pre-trial education atlases and rapid review quality assurance (QA) workflows, broader adoption of IPA sparing in practice or its IPA analysis post-hoc dosimetry studies or other trials will remain limited by such resource intensive measures. Novel tools are needed to shorten learning curves of investigators and to accelerate QA.

Indeed, contouring inconsistencies are known but understudied in clinical radiation therapy trials. Previously, Thor et al., [4] used a coherent heart definition enabled through their open-source deep learning (DL) algorithm to evaluate heart doses in the RTOG 0617 trial. The trial heart doses were found to be significantly higher than previously reported, which they concluded may have led to an even higher mortality rate. Auto-segmentation is likely to reduce contouring and dose inconsistencies while increasing the quality of clinical RT trials and routine clinical treatments.

Automatic segmentation has drawn enormous attention in RT since it reduces the contouring time drastically and creates contours with less intra- and interobserver variability. Recently, convolutional neural networks (CNN) were adopted to advance structure delineation accuracy significantly. Many groups have designed novel deep CNN architectures for pelvic organ segmentation [15-23]. Several of these papers have focused specifically on preoperative prostate GTV and OARs [15-21], and a few on post-operative prostate CTV and OARs [22,23].

Automating IPA segmentation with artificial intelligence (AI) modeling has some challenges: 1). historically labelled data for IPA segmentation is scarce and the available labels are noisy. The IPA was not conventionally considered an organ-at-risk (OAR) and large variation was observed in the starting and ending slice of the IPA contour in the cranial-caudal direction. Furthermore, due to the resolution of contouring tools, it is unavoidable that the IPA contour contain muscle or bones, leading to noise in the training data set. 2). Due to the limited soft tissue contrast in CT images, the IPA is often contoured with the MRI fusion. However, the final registration saved in the treatment planning system is mostly registered to the fiducials for prostate segmentation, which is suboptimal registration for accurately segmenting the IPA. Ideally, the AI model should be able to learn from this offset. 3). MRI is not always available for patients, so the AI model should be flexible regarding input data and be able to accurately segment the IPA with or without the presence of MRI (i.e. CT alone).

In this work, we aimed to address all the above challenges and develop an auto-segmentation model for the IPA to reduce contouring inconsistences, to improve uniformity in IPA segmentation for clinical trials, and to facilitate the introduction of IPA for clinical planning. Allowing model training with a noisy data set is more than needed for future model maintenance when multiple institution data is included. To the best of our knowledge, this is the first work that performs automated IPA segmentation – deep learning or otherwise.

## II. MATERIALS & METHODS

### Dataset

The model training and evaluation dataset consists of CT and T2 weighted MRI for 56 patients with prostate cancer treated at a large medical institution in the United States from 2017 to 2021. In this dataset, the IPA was segmented by multiple physicians and reviewed and revised by an expert physician. All CT images were acquired using a 16-slice CT scanner (Royal Philips Electronics, Eindhoven, The Netherlands), and these CT images were acquired with voxel size 1.17mm×1.17mm×2.5mm. MRI images were acquired with a voxel size of 0.56mm×0.56mm×3mm. The MRI was a rigid registered CT with a fiducial-based registration most appropriate for prostate segmentation. Fourteen patients were used as the test dataset, and the remaining 42 patients were used for cross-validation. The IPA labels were cleaned to ensure uniformity in the extent of the structure to be segmented. The caudal structure limit was determined as 1 cm beyond the primary target (prostate) planning treatment volume (PTV).

An additional 30 patients treated at the same institution from 2020 to 2021 were recruited for two "inexperienced" physicians (no prior IPA segmentation training) to contour with and without the assistance of the AI model as part of the observer study, since this AI model was developed to assist physicians in contouring the IPA. The details of the observer study are described below.

## Model

The model architecture has a UNet [24] backbone. The model was designed to satisfy three conditions: effective learning of small structures, learning appropriate and mutually exclusive features from MRI and CT, learning the effective flow between MRI and CT for an effective combination of features while avoiding additional rigid registration to bones, and to produce acceptable segmentation when MRI is unavailable. The model architecture proposed is shown in Figure 1, and each component of the model is explained in detail below.

### Squeeze and Excite Blocks (SE) to Improve Feature Loss in Downsampling

Standard U-Net contains multiple downsampling layers to facilitate learning of high-level features. Too many downsampling layers can hurt the segmentation of a small structure like the IPA and lose features for model training. We restricted the number of pooling layers in the network to two and included SE blocks to improve the feature learning. The SE [25] layer improved the representational power of the network by enabling it to perform dynamic channel-wise feature recalibration. The 3D SE learning extracted 3D features from the CT image directly by extending two-dimensional squeeze, excitation, scale and convolutional functions to three-dimensional functions.

### Modality Attention to Enhance Feature Learning from MRI and CT

To enable effective learning of features from MRI and CT, we used an effective modality attention block in the skip connection before concatenating them with the decoder layers. Separate encoders were used for extracting MRI and CT features. For designing an effective modality attention unit, several conditions need to be met: 1). It should be completely data-driven, 2). It should be able to perform attention effectively from two raw feature maps that are offset with each other, and 3). It should be light enough to avoid additional computational overhead. To satisfy all three conditions, it is necessary to consider feature-based self-attention instead of any attention mechanism that involves computationally complex dot products. We constructed modality attention using global average pooling and fully connected layers. The modality attention block consisted of a deformable convolution layer [26], which learns the offsets for the MRI necessary for effectively comparing MRI features with that of the CT. Normal convolution layers cannot model large unknown transformations because of their fixed geometric structure. The 3D deformable convolution layer adds a 3D offset to enable learning of a deformed sampling grid. These offsets are learned simultaneously along with convolution parameters.

The MRI and CT features go through the following steps in the modality attention module:

$$Z_{MRI}^{Tran} = DefCon(Z_{MRI}) \tag{1}$$

$$Z_{MRICT} = Z_{CT} + Z_{MRI}^{Tran} \tag{2}$$

$$Z_{GP} = \frac{1}{W \times H \times D} \sum_{i=1}^{W} \sum_{j=1}^{H} \sum_{k=1}^{D} Z_{MRICT}(i,j,k) \tag{3}$$

$$Z_{CT}^{att} = SoftMax(Z_{GP}) \times Z_{CT} \tag{4}$$

$$Z_{MRI}^{att} = SoftMax(Z_{GP}) \times Z_{MRI}^{Tran} \tag{5}$$

$$Z_{encoder} = Z_{CT}^{att} + Z_{MRI}^{att} \tag{6}$$

**Modality Dropout (MoDO) to Allow Flexibility of MRI and/or CT as Inputs**

To ensure that the model can work well with only CT as input, we used MoDO where MRI is dropped out randomly from the input during training. This training strategy forces the network to use information from both modalities, even when one modality is so highly correlated with output that the other would otherwise largely be ignored. This also enables the model to be used in the absence of MRI as input.

## Loss Functions to Correct Bone/Muscles in IPA Training Data

The IPA is a small structure and the contour resolution is limited. It is commonly seen that IPA structures include some muscle and bone tissue, which can lead to noise in the training data and inferior model performance. Therefore, an unsupervised approach is employed to correct the labels. Since MRI has most of the important anatomical features for IPA segmentation, adding an unsupervised approach for cleaning the labels off muscle and bone based on CT forces the model to learn CT features. In this work, we corrected the labels by training the segmentation model using a self-supervised muscle and bone correction loss. The loss function used for training the model is shown in Equation (7).

$$Loss = L_{DSC} + w_m L^{SS}_{Mucle_inclusion} + w_b L^{SS}_{Bone_inclusion} \qquad (7)$$

$$L_{DSC} = 1 - \frac{(Pred \cap GT)}{(Pred \cup GT)} \qquad (8)$$

$$L^{SS}_{Mucle_inclusion} = \frac{(Pred \cap CT_{Muscle})}{(Pred)} \qquad (9)$$

$$L^{SS}_{Bone_inclusion} = \frac{(Pred \cap CT_{Bone})}{(Pred)} \qquad (10)$$

$Pred:\text{Model Prediction}\,;\ GT:\text{Physician Contour};\ CT_{Muscle}:\ -29 \leq CT \leq 150;\ CT_{Bone}:\ 300 \leq CT \leq 1200;\ w_m = 0.1, w_b = 0.01$

Muscle and bone inclusion loss is calculated between the model prediction and muscle/bone threshold CT. $w_m\ and\ w_b$ were optimized using grid search.

## Training Details

For training the model we randomly applied data augmentation techniques such as rotating by small angles (<10°), image scaling, and image flipping for more effective learning. Adaptive histogram equalization was used for data preprocessing to enhance edge definitions and improve local contrast for CT, while min-max normalization was used for MRI. The network was trained with Adam optimizer with an initial learning rate of 0.001 with exponential decay. This was implemented in Keras with TensorFlow backend and trained on one 32GB NVIDIA Tesla V100 GPU. The batch size was four due to memory limitations.

## Model Performance Evaluation

We performed ablation studies to understand the importance of each unique component of the model and to evaluate their effectiveness. The UNet model with a single encoder, squeeze and excite blocks, and without modality attention served as the baseline model in our study and is denoted as 'SNet' in the comparison. The proposed architecture that has separate encoders for MRI, CT, and modality attention is denoted as 'SNet-MA' in the comparison. We compared the DSC values of predicted IPAs by using paired two-sided t-tests for the following network architectures:

(1) SNet – baseline model trained with only DSC loss;

(2) SNet with Muscle-Bone loss – baseline model trained with DSC+Muscle-Bone loss;

(3) SNet-MA – proposed architecture but trained with only DSC loss; and

(4) SNet-MA with Muscle-Bone loss - proposed architecture and trained with DSC+Muscle-Bone loss.

Additionally, we evaluated the effectiveness of modality dropout to allow the flexibility of MRI and/or CT as input for the IPA segmentation. We trained SNet-MA with the Muscle-Bone loss model without and with Modality Dropout (MoDO) and compared the contour quality with CT only as input versus MRI and CT as input.

We reported the segmentation results using three quantitative metrics: Dice similarity coefficient score (DSC) as a percentage, average surface distance (ASD) in millimeters, and Hausdorff95 (HD95) in millimeters. Together, these metrics provided a comprehensive quantitative evaluation.

## Dosimetric Assessment

We further evaluated the dosimetric accuracy of the AI segmented IPA structure. We took the clinical plan generated with the physician's IPA contour in the 14 test cases and mapped the planned dose on the AI segmented IPA contour to compare important dose-volume parameters. For the IPA, important dose-volume parameters were defined based upon constraints from our ongoing multi-center randomized trial to be the mean (Dmean) and the volume receiving >=20 Gy dose (V20). These values were compared between manual segmentation and predicted IPA segmentation for the 14 test patients.

### Clinical Observer Study

Since this AI model was developed to assist physicians for contouring the IPA, we hypothesized that the AI model improves contour acceptance when used as "warm start" for inexperienced physicians to contour the IPA. We evaluated our hypothesis by having inexperienced physicians contouring the IPA with and without AI assistance, and compared efficiency and quality improvements. The recorded time taken to complete the contour with and without AI assistance was used as the efficiency measure. Next, an expert physician scored the contours predicted by AI (AI-raw), inexperienced MD's contour without AI (MD-raw), and inexperienced MD's contour with AI assistance (AI-MD).

Scores from 1-4

1. Unacceptable
2. Acceptable with major corrections
3. Acceptable with minor corrections
4. Acceptable as-is

## III. RESULTS

### Quantitative Metrics

Table 1 summarizes the results of the ablation study with the 14 test cases. Our proposed model, SNet with Modality attention (SNet-MA) and Muscle-Bone loss, has a DSC accuracy of 62% and performs significantly ($p<0.005$) better than other models. The SNet baseline model performs better when Muscle-Bone loss is used in combination with DSC loss, as opposed to training with DSC loss alone. SNet-MA without the use of Muscle-Bone loss performs worse than the SNet baseline model. Intuitively this makes sense as important CT features are related to muscle and bone. For the IPA, the remaining anatomical features depend significantly on MRI alone. Without Muscle-Bone loss, an attention framework will not work effectively since the labels are noisy.

Figure 2 shows the boxplots comparing DSC, ASD, and HD95 for various combinations of AI models and input images for IPA segmentation. The proposed model (SNet-MA with Muscle-Bone loss) trained without modality dropout is denoted as 'MRICT,' while the proposed model trained with modality dropout is denoted as 'MoDO' in the figure. MRICT model performance with CT and MRI as input serves as the baseline for comparison. The performance of the MRICT model with CT only as input (Mean DSC 44%) is significantly worse when compared to a CT model trained with CT only (Mean DSC 54.1%). The MoDO training strategy rectifies this problem, and the corresponding model even performs better than the CT-only model when MRI is dropped from the input (57.2%). The MoDO model, SNet with Modality attention trained with DSC+Muscle-Bone loss and MoDO training strategy, performs the best compared to all models, with a DSC of 62.2%, ASD of 2.54 mm and HD95 of 7mm.

Figure 3 illustrates two example patients with physician and AI segmented IPA contours with CT alone as input versus MRI alone as input. It can be observed that most of the features for segmentation come from MRI. However, AI segmentation with CT alone as input also resulted in satisfactory quality of contour.

**Dosimetric Assessment**

Figure 4 shows the DVH comparison for physician and AI segmented IPA contours for the 14 test cases. Clinical treated plan dose mapped on AI segmented IPA contour demonstrated no difference in V20, and the differences in Dmean were within 1% and statistically and clinically insignificant.

## Clinical Observer Study

Figure 5a shows the boxplot for the time difference when an inexperienced physician started from scratch (MD-raw) and started from AI-predicted IPA (AI-MD). The mean contouring times without and with AI assistance were 10.8 vs 4.7 minutes, respectively, with the difference observed statistically significant. The scores received by the AI-predicted (AI-raw), inexperienced MD starting from scratch (MD-raw) and starting from AI-predicted contour (AI-MD) IPA contours are tabulated in Figure 5b. AI-raw contours without any physician intervention received an average score of 3.7. A perfect score of 4 was received by 26 out of 30 patients, and one patient received a score of 3. There were 3 out of 30 patients with a score of 2 or less. MD-raw contours received an average score of 3.1. A perfect score of 4 was received by 14 out of 30 contours, and 6 out of 30 received a score of 3. Ten patients received a score of 2 or below. AI-MD contours were significantly improved over MD-raw ($p<0.05$), with 23 out of 30 contours receiving a score of 4. The average score received by contours in the AI-MD dataset improved to 3.7. Only one patient received a score of 2 or below.

In Figure 5c, for 55% of the patients, the corrections made by the inexperienced physician in AI-MD contours were deemed unnecessary, as the experienced physician preferred AI-raw over AI-MD contour for these patients. For 31% of the patients, the changes made did not affect the quality of the contour, and for 14% of the patients, the experienced physician preferred the inexperienced physician-corrected AI contour over the AI-raw contour.

# IV. DISCUSSION AND CONCLUSIONS

Approximately 50% of prostate cancer patients could develop radiation-induced erectile dysfunction (RiED). A variety of studies suggest that RiED is mediated by vascular dysfunction at the level of small arteries such as the IPA, which is not segmented or considered an OAR in conventional radiation therapy. This study set out to employ a deep CNN for IPA segmentation, with the aim of making the model practical for clinical use and maintenance.

To address the label variability problem, we proposed an innovative loss function that reduces the error in prediction in an unsupervised manner, and additionally leverages the trial protocol for uniformity in the extent of contouring the IPA. To learn from the imperfectly registered MRI and CT, we designed a modality attention framework that uses deformable convolutions to learn the offset with a lightweight attention mechanism for effective feature learning. We trained the model with a modality dropout strategy that enables the model to perform well, even without MRI. We evaluated the geometric and dosimetric accuracy of the AI predicted contour and also quantified the improvement of contour consistency by having an expert physician reviewing the IPA contours from inexperienced physicians with and without AI assistance. These innovations significantly reduced the manual effort to create a 'perfect' training data set and to make model creation or maintenance more clinically practical.

The modality dropout training strategy translates the information learned from MRI and CT to CT alone, enabling the model to be used without MRI, given that MRI may not be available for all patients. Even though the model output is highly correlated with the features in MRI (Figure 3), the proposed model works similar or even slightly better than a model trained to accept only CT in the absence of MRI. In comparison, without modality dropout, dropping out MRI from input significantly degrades the model performance as seen in the 'MRICT' model performance in Figure 2.

The AI segmented IPA was similar to manual contours delineated by expert oncologists. Considering the small volume of the IPA contour, any small discrepancy in the contour can lead to significant deduction of the metric for contour quality evaluation. One limitation of this work is the lack of a baseline on 'satisfactory' DSC score for IPA contour quality. Hence, we utilized the dosimetric accuracy to demonstrate that the AI segmented IPA contours are dosimetrically equivalent to the expert physician's contour.

The observer study supported the finding that AI contour quality is clinically satisfactory compared to the inexperienced physicians' manual contour, and that it can improve uniformity and efficiency. It is also evident that AI assistance improved the inexperienced physicians' contour score (Figure 5).

Deep learning models when trained with only a single institution data sometimes exhibit a generalizability issue when it comes to medical image segmentation. This is a limitation of this study and further testing on external datasets if they become available in the future would be beneficial. Furthermore, deep learning models benefit from large datasets but because of IPA being a rare structure the amount of data available is small. Other limitations of this study include reliance on a single "expert" reviewer of contours, and lack of validation of IPA dose constraint relationships to ED outcomes.

In summary, an AI-based IPA segmentation model is developed in this study. There are several potential applications of the developed model: 1) quality assurance of IPA contours for clinical trials; 2) leveraging AI to assist inexperienced physicians for contouring IPA; 3) automatically generated IPA contours for use in routine clinical planning to collect large-scale IPA dosimetry data with minimal additional effort that can benefit the toxicity and outcomes of study; 4) automated generation of IPA contours in existing datasets of already treated patients to correlate to outcomes with minimal effort compared to traditional manual investigator contours subject to biases.

**Acknowledgement**

We would like to thank Varian Medical Systems Inc. for supporting this study and Ms. Sepeadeh Radpour for editing the manuscript.